# A Unified Glassy Rheology for Granular Matter

Zhikun Zeng[1,*], Jiazhao Xu[1,*], Hanyu Li[2,*], Shiang Zhang[3], Houfei Yuan[1], Chijin Zhou[2],

Xueliang Dai[2], Haiyang Lu[1], Xin Wang[1], Jun Zhao[3], Yonglun Jiang[2], Zhuan Ge[2],

Gang Huang[2], Chengjie Xia[4], Jianqi Sun[3,†], Yan Xi[5,‡], and Yujie Wang[1,2,6,§]

[1]*School of Physics and Astronomy, Shanghai Jiao Tong University, Shanghai, China.*
[2]*College of Physics, Chengdu University of Technology, Chengdu, China.*
[3]*Med-X Research Institute and School of Biomedical Engineering, Shanghai Jiao Tong University, Shanghai, China.*
[4]*School of Physics, East China Normal University, Shanghai, China.*
[5]*Jiangsu First-Imaging Medical Equipment Co., Ltd., Jiangsu, China.*
[6]*State Key Laboratory of Geohazard Prevention and Geoenvironment Protection, Chengdu University of Technology, Chengdu, China.*

[*]These authors contributed equally to this work.
[†]Contact author: milesun@sjtu.edu.cn
[‡]Contact author: yanxi@first-imaging.com
[§]Contact author: yujiewang@sjtu.edu.cn

Granular flows are ubiquitous in nature and industrial applications, yet a complete continuum theory remains a long-standing challenge. The leading empirical approach, $\mu(I)$ rheology, lacks microscopic foundations and becomes multivalued in dense, slowly sheared flows where nonlocal corrections are required. Exploiting state-of-the-art high-speed X-ray tomography to investigate microscopic dynamics of dense granular flows in a Couette geometry, we establish a new, universal constitutive law spanning quasi-static to inertial regimes based on structural relaxation, resolving the fundamental difficulty in the original $\mu(I)$ framework. By further establishing a non-equilibrium statistical framework for granular flows, we demonstrate an intrinsic analogy between driven granular matter and hard-sphere liquids owing to their identical Carnahan-Starling equation of state, naturally explaining our rheological approach and the emergence of glassy behaviors. Our framework unifies granular rheology with the broader physics of disordered systems and provides a complete, microscopically-based theoretical framework for dense granular flow.

## I. INTRODUCTION

Granular flows play a central role in natural and industrial processes [1-4], from geophysical hazards to industrial powder handling. Although granular materials can flow like liquids, their flow behavior differs fundamentally from that of ordinary fluids and spans a rich spectrum of phenomena, including multiple flow states and their coexistence under varying external driving [5,6], shear localization [7,8], hysteresis [9,10], creep [11,12], secondary rheology [13], and flow thresholds that depend on system size [14]. These phenomena arise from the intrinsic nature of granular matter as being athermal, disordered, nonlinear, multiscale, and lacking clear scale separation [15,16]. To date, no continuum theory grounded in a clear microscopic understanding has succeeded in developing a unified constitutive law to capture these diverse behaviors in granular flows.

Currently, the most widely used constitutive description for granular flow is the empirical $\mu(I)$ rheology [16,17]. It states that the macroscopic friction coefficient $\mu$, defined as the ratio between shear stress $\sigma$ and pressure $P$, depends on a single dimensionless inertial number $I = \frac{\dot{\gamma}d}{\sqrt{P/\rho_s}}$, which is essentially the ratio between the imposed shear timescale and an empirically defined, pressure-controlled microscopic timescale. While $\mu(I)$ successfully captures homogeneous flows at moderate densities, it breaks down in dense flows where hysteresis, secondary rheology, and finite-size effects arise, rendering the constitutive relation multivalued

[18]. This breakdown arises because particles in dense granular flows interact not only through collisions but also via persistent frictional contacts, which introduce unique microscopic pathways of structural relaxation and energy dissipation that differ fundamentally from the pressure-accelerated relaxation assumed in the $\mu(I)$ framework [19]. To overcome these limitations, a nonlocal extension of $\mu(I)$ rheology has been developed [20]. By adapting the kinetic-elasto-plastic (KEP) framework from soft glassy materials [21,22], this approach introduces a spatial cooperativity length through a nonlocal fluidity equation that couples local flow to its surroundings. Although this approach successfully reproduces a broad range of flow geometries, it remains phenomenological, as the spatial cooperativity is introduced empirically and lacks a clear microscopic understanding [23]. In this context, a predictive and unified theory for dense granular flows requires the development of a constitutive framework grounded in a thorough understanding of the underlying microscopic processes that are directly accessible in experiments, which remains lacking to date.

To address these challenges, we develop a state-of-the-art high-speed X-ray tomographic system that, for the first time, enables three-dimensional (3D), particle-resolved imaging of dense granular flows, capturing both the dynamically evolving packing structure and the complete trajectories of every grain with spatial accuracy comparable to static tomography. By replacing the empirically defined timescale in the original $\mu(I)$ relation with the true structural relaxation time, we establish a linear and local constitutive relation that unifies granular flow from quasi-static to inertial regimes without invoking nonlocal corrections. Extending the Edwards ensemble, the non-equilibrium statistical mechanical framework originally developed for static granular packings, to flowing states, we further show that granular flows obey the hard-sphere liquid equation of state, thereby establishing an intrinsic analogy between dense granular flows and disordered thermal systems. This correspondence provides a theoretical foundation for our rheological approach: the new non-equilibrium statistical mechanics and liquid-state theoretical approach properly captures the coupled collisional and frictional dynamics underlying the traversal of a complex phase space and clarifies the glassy nature of complex phenomena associated with dense granular flow. Together, we establish a general and predictive constitutive framework for granular flows rooted in a thorough understanding of the underlying microscopic processes.

## II. RESULTS

### A. Experimental test of local $\mu(I)$ rheology

Establishing a microscopic foundation for granular rheology requires direct experimental access to time-resolved particle-scale dynamics inside 3D granular flows. To address this challenge, we developed a high-speed X-ray tomographic system comprising 29 source-detector pairs arranged around a granular shear cell [Fig. 1(a)] and a neural-network-assisted reconstruction algorithm (see Appendix A for details). With 29 sources operating in parallel, 3D imaging is achieved at 30 Hz with a spatial resolution of 160 μm, allowing full tracking of all particles across a wide range of flow conditions. Experiments were performed in an annular Couette shear cell, where approximately 9,000 monodisperse plastic spheres of diameter $d = 3$ mm were sheared by rotating the inner cylinder at different constant rotation rates $\Omega = 0.01 - 6$ rps under controlled normal pressure $P \approx 1500$ Pa [Fig. 1(b)]. The torque on the inner cylinder was recorded simultaneously, allowing determination of the radial shear-stress profile across the flow. In parallel, dynamic X-ray imaging was carried out with our tomographic system to obtain particle-resolved 3D configurations in real time during shear. Particle centroid coordinates and trajectories were extracted, from which local volume fractions $\phi$ [the inset of Fig. 1(c)] and velocity fields were obtained. We conducted two sets of experiments using particles with different interparticle friction: smooth-surface (SS) and rough-surface (RS) grains. Details are provided in the *Methods* section.

Owing to the cylindrical symmetry of the steady annular Couette flow, we divide the annulus into concentric rings of equal width ($0.6d$), and then average the packing structure and particle motions within each ring at different radial positions $R$. Accordingly, the inertial number $I$ and macroscopic friction coefficient $\mu$ at each $R$ are deduced from the local velocity and stress fields (see *Methods*). These measurements enable a direct test of the local $\mu(I)$ rheology [red dashed curve in Fig. 1(c)],

$$\mu(I) = \mu_s + \frac{\mu_m - \mu_s}{1 + I_0/I}, \tag{1}$$

where $\mu_s$, $\mu_m$ and $I_0$ are standard fitting parameters [17]. The measured $\mu$ increases monotonically with $I$, as expected from the $\mu(I)$ relation, but these $\mu(I)$ curves depend strongly on the driving rate $\Omega$ and exhibit clear multivalued behaviors especially at low $I$. In particular, $\mu$ is significantly larger than the $\mu(I)$ prediction at low $\Omega$, while steady flow persists even when $\mu$ falls below the expected yield value $\mu_s$, behaviors that are well-known and have previously been interpreted as requiring finite-size or nonlocal corrections in confined granular flows [20,24]. These observations demonstrate that a simple local $\mu(I)$ law is insufficient to fully capture the rheology of dense granular flow, motivating a microscopic reexamination of its physical basis.

## B. Microscopic timescales for different flow regimes

To understand the origin of the limitations of the local $\mu(I)$ rheology, we investigate the microscopic particle dynamics underlying the observed macroscopic flow by measuring the mean-squared displacement (MSD), $\langle \Delta z^2(\Delta t)\rangle$, along the vertical direction at different $R$, a direction orthogonal to the shear direction and with negligible mean flow [Fig. 2(a); see *Methods* for more details]. The MSDs reveal strong radial variations, with particle motion slowing down rapidly away from the shearing inner boundary. Investigation of particle trajectories further indicates a crossover from collision-dominated to contact-dominated flow regimes as local shear rate decreases. Video 1 shows particle dynamics under shear, and Supplemental Material Fig. S3 shows representative trajectories [25].

We extract two characteristic timescales from the MSDs across different flow regimes. The first is a diffusive timescale defined at the short-time end of the MSD: $\tau_D = \frac{\Delta t \cdot d}{\sqrt{\langle \Delta z^2(\Delta t)\rangle}}$, corresponding to the time required for a particle to diffuse a distance $d$. $\tau_D$ characterizes fast local particle motions, analogous to the Brownian timescale in colloidal systems [26]. While the value of $\tau_D$ slightly depends on the chosen observation interval $\Delta t$, we have verified that it hardly affects the conclusions (*Methods*). Notably, $\tau_D$ itself is driving-dependent across various flow regimes [Fig. 2(c)], reflecting the dissipative nature of granular matter, in which microscopic dynamics under steady shear is determined by a dynamic balance between energy input by shear and dissipation through inelastic particle collisions and frictional contacts [27]. The second timescale $\tau_F$ is a standard measure of structural relaxation in liquid-state theory defined from the self-intermediate scattering function $F_s(\Delta t)$ [Fig. 2(b); see *Methods*]. $\tau_F$ is obtained by fitting $F_s$ with a Kohlrausch-Williams-Watts (KWW) form, $F_s(\Delta t) = \exp[-(\Delta t/\tau_F)^\beta]$, which also yields the stretching exponent $\beta$ characterizing the heterogeneity of relaxation.

Figure 2(d) shows the ratio of the two timescales, $\tau_F/\tau_D$, as a function of the local volume fraction $\phi$. In the dilute, fast-shear regime ($\phi < 0.56$), structural relaxation is governed by collision-dominated diffusion, such that $\tau_F \approx \tau_D$; in contrast, the two timescales deviate significantly at higher densities ($\phi > 0.56$). This is reminiscent of the familiar decoupling between local caged motion and structural relaxation in supercooled liquids [28]. This glass-like behavior is further evidenced by the evolution of the stretching exponent $\beta$ from the KWW fits. $\beta$ remains near unity for $\phi < 0.56$ but decreases to ~0.3 at larger $\phi$, implying increasingly heterogeneous relaxation involving multiple mechanisms [the inset of Fig. 2(d)]. Moreover, the ratio $\tau_F/\tau_D$ grows rapidly with $\phi$ and diverges in a Vogel-Fulcher-Tammann-like (VFT-like) manner, characteristic of a glassy system. These results demonstrate that the dynamics of driven granular matter depend not only on the balance between energy input and dissipation, but are also sensitively governed by the emergence of glassy dynamics, questioning the oversimplified assumption of the microscopic timescale underlying the $\mu(I)$ framework [29].

### C. Unified glassy rheology for granular flow

Once we determine the relevant microscopic timescales ($\tau_D$ and $\tau_F$) from experiments, we reformulate the $\mu(I)$ rheology by replacing its empirical microscopic timescale $t_{\text{micro}} = \frac{d}{\sqrt{P/\rho_s}}$ with them. When $\mu$ is plotted against the bare Péclet number [30], $\text{Pe} = \tau_D/\dot{\gamma}^{-1}$, datasets that are otherwise separated in ordinary $\mu(I)$ representation collapse onto a single universal curve spanning all flow regimes [see Fig. 3(a)]. Interestingly, the collapsed data follow a Herschel-Bulkley form, a hallmark of viscoplastic fluid rheology widely observed in glassy soft-matter systems such as colloidal suspensions and emulsions [31]. To understand the origin of this Herschel-Bulkley form, we replace the diffusive timescale $\tau_D$ with the structural relaxation time $\tau_F$ to define the Weissenberg number [32], Wi = $\tau_F/\dot{\gamma}^{-1}$ (equivalently, the Deborah number commonly used in viscoelastic models), and examine its relationship with $\mu$. In this representation, all data collapse again onto a single, nearly linear $\mu$(Wi) relation [Fig. 3(b)]. This linearity reveals that macroscopic flow response is fundamentally governed by the competition between the timescale of structural rearrangements that relax stresses and that of externally imposed deformation, mirroring the Maxwell-type description of liquids [33,34].

We proceed to clarify the physical origin of the multivalued behavior in local $\mu(I)$ rheology. At large $I$, both $\tau_D$ and $\tau_F$ are determined by particle collisions and thus carry similar physical content to $t_{\text{micro}}$ underlying the $\mu(I)$ framework. This correspondence ensures a single-valued relation between these timescales and $t_{\text{micro}}$ (and hence $I$). Meanwhile, in dense and slowly sheared regimes, enduring contacts arise alongside inelastic collisions. Previous work has shown that significant creep dynamics exists even when particles maintain permanent contacts in quasi-static shear experiments [19]. Microscopic dynamics in this regime are therefore mediated by both collisional and contact processes, which no longer admit a simple, unified dependence solely on the pressure $P$. Consequently, the single pressure-based $t_{\text{micro}}$ ceases to be useful, leading to the observed multi-valued $\mu(I)$ behavior in dense flows. From this perspective, $\mu(I)$ rheology emerges as a limiting case of a more general, microscopically grounded rheological framework, valid only when microscopic relaxation remains solely collisional and weakly dependent on shear rate.

### D. Hard-sphere glass nature of dense granular flow

A predictive theoretical description of dense granular flow requires a framework that correctly accounts for the coupled collisional and frictional dissipation and captures the emergence of slow, glassy relaxation intrinsic to granular matter. Given the analogous microscopic dynamics between driven granular flow and thermal hard-sphere fluids, we draw on the liquid-state theoretical framework developed for thermal hard-sphere systems. In thermal hard-sphere systems, increasing volume fraction leads to dramatic slowing down of structural relaxation, and transport properties are no longer governed by the equilibrium kinetic temperature $T_k$ but instead by an effective temperature $T_{\text{eff}}$ that characterizes slow structural relaxation in the complex glassy energy landscape [35,36]. In the following, we show that an analogous framework can be established for driven granular fluids, which properly captures the coupled collisional and frictional dynamics underlying the traversal of a complex phase space.

Granular materials are inherently athermal systems whose macroscopic properties are not governed by thermodynamic temperature. Nevertheless, two temperature-like quantities are commonly used: a kinetic temperature $T_k$ for low-$\phi$ granular gases and an Edwards temperature $T_{\text{EDW}}$ for static or slowly sheared high-$\phi$ granular packings. In the granular gas regime, analogous to an ideal gas, a granular kinetic temperature can be defined from the mean kinetic energy per particle [37], $T_k = \langle m \delta v_i^2 \rangle$, where $\delta v_i = v_i - \langle v_i \rangle$ denotes the velocity fluctuation along direction $i$ (here $i = z$, consistent with the MSD analysis). For granular packings, where particles remain static and $T_k = 0$, the Edwards volume ensemble provides instead the appropriate statistical framework, with the configurational temperature $T_{\text{EDW}}$ serving as the relevant temperature-like variable [38]. $T_{\text{EDW}}$ essentially characterizes statistical sampling among mechanically stable configurations of static or quasi-statically driven granular materials, analogous to the effective temperature $T_{\text{eff}}$ in glass physics that governs sampling of inherent states in the glass landscape [35,39,40]. The difference between the two lies in that $T_{\text{eff}}$ accounts for not only the configurational contributions but also for kinetic ones. We therefore generalize the Edwards ensemble to dense granular flows by relaxing the strict mechanical stability constraint to include both contributions [41]. This leads to an effective temperature defined as: $\frac{1}{T_{\text{eff}}} = \frac{1}{T_{\text{EDW}}} + \frac{1}{T_{k,\text{RLP}}}$, which incorporates contributions from both $T_k$ and $T_{\text{EDW}}$ within a unified statistical framework. This generalization preserves ergodicity, as assumed by the Edwards volume ensemble, in the presence of collisions and frictional contacts by retaining the Boltzmann-like distribution underlying the exploration of the complex phase space. The kinetic term corresponds simply to a redefinition of the reference temperature, maintaining the original statistical framework unchanged and self-consistent (see Appendix B for details). As shown in Fig. 4(a), $T_{\text{eff}}$ and $T_k$ coincide in dilute steady flows ($\phi < \phi_{\text{RLP}}$), where short-time motion and configurational relaxation have not yet separated. Beyond the RLP density $\phi_{\text{RLP}}$, $T_{\text{eff}}$ progressively decouples from $T_k$, revealing a clear violation of the Stokes-Einstein relation as in glassy systems.

Critically, this new effective temperature $T_{\text{eff}}$ yields an equation of state identical to that of hard-sphere liquids. As shown in Fig. 4(b), the pressure $P$, volume fraction $\phi$, and $T_{\text{eff}}$ collapse onto a single relation across all experimental conditions. This equation of state follows the Carnahan-Starling form for hard-sphere liquids across the entire range of $\phi$ investigated, extending well into the deep supercooled regime when expressed in terms of $T_{\text{eff}}$,

$$\frac{P}{T_{\text{eff}}} = \frac{\phi}{\pi d^3/6} \cdot \frac{1 + \phi + \phi^2 - \phi^3}{(1-\phi)^3}. \tag{2}$$

The emergence of a hard-sphere equation of state suggests that dense granular matter behaves, by nature, very similarly to supercooled hard-sphere liquids [42,43], despite features absent in thermal hard-sphere liquids such as frictional interactions and complex relaxation dynamics of granular systems at high $\phi$. Hence, $T_{\text{eff}}$ can be identified as the true effective temperature governing the sampling of phase space in dense granular flows.

### E. Relationship with nonlocal models

Based on the newly established equation of state for granular materials, our results directly lead to a state relation between the macroscopic friction coefficient $\mu$ and the packing fraction $\phi$, as shown in Fig. 5(a), reminiscent of the viscosity-$\phi$ relation in supercooled hard-sphere liquids [44]. For each particle type, data obtained over a wide range of driving conditions collapse onto a simple proportional relationship: $(\mu - \mu_s) \propto (\phi_{\text{RLP}} - \phi)$, confirming that the rheological response is solely governed by a measurable state variable $\phi$. Material parameters such as the particle restitution and friction coefficient could potentially alter the proportionality constant. Notably, for two systems with different surface properties, the static yield friction $\mu_s$ (0.43 for SS and 0.47 for RS), defined in the classical $\mu(I)$ as the onset of steady homogeneous granular flow, coincides with the RLP density $\phi_{\text{RLP}}$ (0.59 for SS and 0.57 for RS), which has been identified in recent work as the onset of glassy behaviors in driven granular systems [45]. This correspondence provides further evidence of our glass-physics approach to granular flows. It also indicates that sub-yield creep in dense granular materials is by nature a manifestation of glassy structural relaxation once the system starts to explore the glass landscape.

An alternative explanation for the sub-yield creep invokes nonlocal corrections to $\mu(I)$ rheology [20], which introduce an additional continuum equation to derive the spatial distribution of flow by considering the diffusion of the kinetic temperature $T_k$ ("mechanical noise") from neighboring flowing regions [23,46]. This spatial coupling can sustain flow in regions that would otherwise remain static under a purely local rheology, i.e., for $\mu < \mu_s$ ($\phi > \phi_{\text{RLP}}$). However, our results clearly show that, in this regime, $T_k$ is vanishingly small and cannot account for the observed flow. Therefore, an exceedingly long cooperativity length scale representing the nonlocal effects has to be introduced to reproduce the flow that actually happens. Within the present framework, structural relaxation is governed by the effective temperature $T_{\text{eff}}$, and sub-yield flow arises from the spatial diffusion of $T_{\text{eff}}$ rather than $T_k$ [47,48]. Since $T_{\text{eff}}$ does not vary drastically between flowing and static regimes [see the inset of Fig. 5(b)], a constant thermal diffusivity suffices to generate the observed flow profiles [46]. This picture provides a simpler mechanism for sub-yield flow. Interestingly, we also observe a clear maximum in the dynamical correlation length $\xi_C$ at $\phi_{\text{RLP}}$ (or equivalently at $\mu_s$) from the spatial cooperativity of particle dynamics [Fig. 5(b); see Appendix C for details], which can be viewed as a dynamical residue of the zero-temperature jamming transition [49].

## III. DISCUSSION

Using pioneering high-speed X-ray tomography, we establish a unified rheological law for dense granular flow. The key advance lies in replacing the empirically defined microscopic timescale in $\mu(I)$ rheology with the true structural relaxation time that governs stress relaxation.

The validity of this approach is rooted in the fundamental correspondence we establish between granular flow and hard-sphere liquids, which provides a physical basis for the similar Maxwell-type rheological responses observed in granular flow. Viewed from this correspondence, the microscopic structural relaxation timescale, and hence transport coefficients such as macroscopic friction and thermal diffusivity, can, in principle, be derived from theoretical approaches developed for hard-sphere systems, including Langevin-based descriptions and mode-coupling theories rooted in standard liquid-state theory [50,51], once energy dissipation is appropriately incorporated [52]. This offers a complete closure for a continuum description of dense granular flows within the unified language of non-equilibrium statistical physics and liquid-state theory, representing a significant advance in the understanding of driven, dissipative materials.

## IV. METHODS

### A. Multi-source X-ray tomographic system

Three-dimensional imaging was performed using an in-house developed multi-source X-ray tomographic system comprising 29 fixed X-ray imaging chains uniformly arranged on a circular ring centered on the sample. Each chain consisted of an X-ray source (Suzhou Powersite Electric Co., Ltd., PSM-C5) operated at 100 kV and 30 mA with an exposure time of 5 ms per projection, coupled to a flat-panel detector (IRay Technology, Jupi0606X1) comprising a 768×768 pixel array with a pixel size of 200 μm. A lead collimator mounted at the exit of each source confined the X-ray cone beam and prevented cross-illumination of neighboring chains. All chains were synchronized via an internal Controller Area Network

(CAN) bus for simultaneous acquisition. Projection intensities were calibrated across chains to ensure consistent brightness across all views. The system provided a cylindrical field of view of approximately 8 cm in diameter and 8 cm in height, with a calibrated voxel size of 160 μm, sufficient to fully resolve all 3-mm-diameter particles.

The system was operated in two acquisition modes. In static-imaging mode, a mechanically stable packing, together with the shear cell, was rotated over 180° (180 s) at $1° \cdot s^{-1}$, while projections were acquired at 1 Hz, yielding 29×180 projections. The granular assembly remained stationary relative to the shear cell during the rotation, thereby sampling a fixed packing structure over multiple viewing angles. These dense-view projections were reconstructed using the Feldkamp-Davis-Kress (FDK) algorithm to obtain high-fidelity reference volumes [53]. In dynamic-imaging mode for time-resolved measurements, the shear cell remained stationary while shear was applied internally, and all 29 imaging chains acquired projections simultaneously at 30 Hz. These 29-view projections were reconstructed using a deep-learning-assisted sparse-view pipeline. More details are described in Appendix A.

### B. Annular Couette shear experiments

Monodisperse 3D-printed plastic spheres of diameter $d = 3$ mm and solid density $\rho_s = 1.16 \times 10^3$ kg$\cdot$m$^{-3}$ were used in all experiments. To vary interparticle friction, we prepared smooth-surface (SS) and rough-surface (RS) particles. RS particles with a higher friction coefficient were designed by uniformly decorating each sphere with 400 hemispherical bumps of diameter $0.05d$. The corresponding static friction coefficients were $\mu_{SS} = 0.43$ and $\mu_{RS} = 0.47$, measured by repose-angle tests [54].

Granular flows were generated in an annular Couette shear cell with outer radius $R_1 = 3.4$ cm $(\sim 11.3d)$ and height $H = 6.6$ cm $(\sim 22d)$. The inner cylinder of radius $R_0 = 1.0$ cm $(\sim 3.3d)$ was driven from below by a motorized rotation stage (SURUGA SEIKI CO. LTD., KS402-180). Approximately 9,000 particles were accommodated in the cell. A torque sensor (DAYSENSOR DYN200, maximum measurement range 0.1 N·m, accuracy 0.0001 N·m) mounted between the rotation stage and the inner cylinder recorded the torque $\Gamma_0$ at 1000 Hz. To maintain an approximately constant normal pressure and reduce vertical pressure gradients, a loading mass of 0.5 kg (approximately three times the total particle weight) that translated freely in the vertical direction was placed on top of the packing, corresponding to an applied pressure of $P \approx 1500$ Pa. To reduce boundary slip and suppress crystallization, a monolayer of hemispherical particles was glued to the inner surface of the shear cell.

To access different flow regimes, the inner cylinder was rotated at driving rates $\Omega$ ranging from 0.01 to 6 revolutions per second (rps). Before X-ray acquisition, the system was pre-sheared continuously for 2 min to reach a steady state, identified by the torque $\Gamma_0$ fluctuating around a constant mean value. Time-resolved measurements were then performed at 30 Hz for 2 s. Each experimental condition was repeated for three independent realizations. The profiles of both azimuthal velocity and packing fraction exhibited no systematic dependence on vertical position, allowing all quantities to be averaged over the full cell height.

### C. Particle segmentation and tracking

After reconstruction, particles were segmented as connected voxel clusters using marker-based watershed segmentation implemented in MATLAB [40,55]. Particle positions were obtained from the centroids (centers of mass) of the segmented clusters, and particle volumes from their voxel counts, yielding a centroid uncertainty of approximately $(3 \times 10^{-3})d$. Particle trajectories were extracted using an improved Hungarian tracking method that globally optimizes nearest-neighbor matching between consecutive frames [56], with a maximum allowed displacement of $4d$ between adjacent frames.

### D. Local rheological variables and test of μ(I) rheology

Local shear rate and stress fields were obtained based on the particle trajectories. Owing to azimuthal symmetry under steady shear, the annulus was divided into concentric rings of width $0.6d$ and quantities were averaged within each ring at radial position $R$. The mean azimuthal velocity profile $v_t(R)$ was obtained by averaging particle azimuthal velocities inside each ring. As shown in Supplemental Material Fig. S2(a) [25], the velocity profile $v_t(R)$ exhibits a narrow shear band approximately $4d$ wide near the inner wall. To reduce numerical noise in spatial derivatives, the velocity profile was fitted with an exponential of a fourth-order polynomial [57], $v_t(R) = v_0 \exp(\sum_{n=0}^{4} a_n R^n)$, where $v_0$ and $a_n$ are fitting coefficients, see Supplemental Material Fig. S2(b) [25]. The corresponding local shear rate in cylindrical coordinates was calculated as $\dot{\gamma}(R) = \frac{1}{2}\left(\frac{\partial v_t}{\partial R} - \frac{v_t}{R}\right)$, from which the local inertial number $I(R) = \frac{\dot{\gamma}(R)d}{\sqrt{P/\rho_s}}$ was deduced. The shear stress followed from the torque balance condition as $\sigma(R) = \frac{\Gamma_0}{2\pi H R^2}$, leading to the radial distribution of the macroscopic friction coefficient, $\mu(R) = \frac{\sigma(R)}{P}$. These measurements enable a direct test of the empirical $\mu(I)$ rheology.

### E. Microscopic dynamics

### 1. Mean-squared displacement

We characterized the local dynamics by measuring the mean-squared displacement (MSD) along the vertical $z$-direction at different radial positions $R$:

$$\langle \Delta z^2(R, \Delta t)\rangle = \langle [z_j(R, t_0 + \Delta t) - z_j(R, t_0)]^2 \rangle, \tag{3}$$

where $z_j(R, t)$ is the vertical position of particle $j$ at time $t$, and $\langle \cdots \rangle$ denotes averaging over different particles and starting time $t_0$.

### 2. Diffusive timescale

The diffusive timescale was defined as $\tau_D = \frac{\Delta t \cdot d}{\sqrt{\langle \Delta z^2(\Delta t)\rangle}}$, which represents the time required for a particle to diffuse a distance of one grain diameter $d$. This operational definition captures short-time particle motion, which would ideally be extracted from the ballistic regime of the MSDs but cannot be directly resolved here owing to limited temporal resolution. In the dilute, fast-shear regime ($\phi < 0.56$), $\tau_D$ is expected to be comparable to the structural relaxation time $\tau_F$, as structural relaxation is solely governed by collision-dominated diffusion. This relation provides a basis for calibrating the observation interval to $\Delta t = 1/24$ s. We verified that varying $\Delta t$ within the short-time window does not affect any qualitative trends or conclusions.

### 3. Self-intermediate scattering function and structural relaxation

Structural relaxation was characterized by the self-intermediate scattering function along the $z$-direction, $F_s(q, \Delta t) = \langle e^{-iq \cdot \Delta z(\Delta t)} \rangle$, with $q = 3.5/d$ corresponding to the first peak of the static structure factor, and $\langle \cdots \rangle$ denotes averaging over different particles and starting time

$t_0$.

## ACKNOWLEDGEMENTS

We thank W. Kob for helpful discussions. The work has been supported by the National Natural Science Foundation of China (No. 12534008, 12274292, 62471297, 12474193) and the Space Application System of China Manned Space Program (No. KJZ-YY-NLT0504). Z.Z. acknowledges support from the National Natural Science Foundation of China (No. 123B2060).

## APPENDIX A: DEEP-LEARNING-ASSISTED SPARSE-VIEW RECONSTRUCTION

Volumetric reconstructions from sparse-view projections were obtained using a deep-learning-assisted reconstruction pipeline. For each time frame, an initial 3D volume was reconstructed from the 29-view projection set using the FDK algorithm, as shown in Supplemental Material Fig. S1(a) [25]. Owing to the ultra-sparse angular sampling, these initial volumes contained pronounced streak artifacts. We then refined the artifact-corrupted volumes using the Sliding-volume Streak Artifact Reduction Network (S-STAR Net) [58], which suppresses streak artifacts while preserving particle morphology, as shown in the Supplemental Material Fig. S1(b) [25], enabling reliable segmentation and tracking.

S-STAR Net formulates sparse-view reconstruction as a domain-translation task between an ultra-sparse-view domain $U$ and a high-quality full-view domain $F$, as shown in the Supplemental Material Fig. S1(c) [25]. The full-view domain $F$ was obtained from static-imaging mode reconstructions. The network learns a forward mapping $A: U \to F$ and a reverse mapping $B: F \to U$, with cycle consistency enforced by $B[A(u)] \approx u$ for $u \in U$. Two discriminators, $D_F$ and $D_U$, constrain the translated volumes to match the statistical distributions of domains $F$ and $U$, respectively.

To exploit the 3D continuity of the CT data, S-STAR Net operates on 3D sub-volumes (128 × 128 × 24 pixels) rather than individual 2D slices. During training, sub-volumes were randomly sampled from reconstructed volumes to increase data diversity. At inference, sub-volumes were processed sequentially according to their spatial positions and stitched to form the final 3D volume after being processed by S-STAR Net. This approach improves the discrimination between physical structures and streak artifacts. Each generator in S-STAR Net adopts an encoder-bottleneck-decoder architecture, as shown in the Supplemental Material Fig. S1(d) [25], with the bottleneck composed of volume-attention residual (VAR) blocks that compute cross-dimensional attention across channels and spatial dimensions (depth, height, and width). As shown in the Supplemental Material Fig. S1(e) [25], the operation of a VAR block on an input tensor $x$ can be written compactly as:

$$\text{VAR}(x) = x + W_{\text{conv}}(x) + W_{\text{att}}(x), \tag{A1}$$

where $W_{\text{conv}}(x)$ denotes the convolutional branch and $W_{\text{att}}(x)$ denotes the volume-attention operator.

The network is trained using a combination of adversarial loss, difference-enhancement (DE) loss, and reconstruction loss. The adversarial loss constrains the generators to produce outputs whose statistical distributions match those of the full-view and sparse-view domains:

$$\mathcal{L}_{\text{GAN}}(A, D_F) = \mathbb{E}[\log D_F(f)] + \mathbb{E}\left[\log\left(1 - D_F(A(u))\right)\right], \tag{A2}$$

$$\mathcal{L}_{\text{GAN}}(B, D_U) = \mathbb{E}[\log D_U(u)] + \mathbb{E}\left[\log\left(1 - D_U(B(f))\right)\right]. \tag{A3}$$

To prevent the generators from introducing artificial structures, the DE loss imposes cycle

consistency by penalizing deviations between an input volume and its cyclically transformed version:

$$\mathcal{L}_{\text{DE}} = \mathbb{E}\left[\|B(A(u)) - u\|_{\text{m}}\right] + \mathbb{E}\left[\|A(B(f)) - f\|_{\text{m}}\right]. \tag{A4}$$

The reconstruction loss directly aligns sparse-view reconstructions with their corresponding full-view references:

$$\mathcal{L}_{\text{rec}} = \mathbb{E}[\|A(u) - f\|_{\text{m}}] + \mathbb{E}[\|B(f) - u\|_{\text{m}}]. \tag{A5}$$

Here $\|\cdot\|_m$ denotes a metric combining the $l_1$ norm, which measures pixel-wise intensity differences, and structural similarity index (SSIM), which preserves local structural contrast. The total loss is given by

$$\mathcal{L}_{\text{total}} = \mathcal{L}_{\text{GAN}}(A, D_F) + \mathcal{L}_{\text{GAN}}(B, D_U) + \lambda_1 \mathcal{L}_{\text{DE}} + \lambda_2 \mathcal{L}_{\text{rec}}, \tag{A6}$$

where $\lambda_1$ and $\lambda_2$ weight the consistency and reconstruction terms. This reconstruction framework yields artifact-suppressed 3D volumes that closely match the experimentally obtained full-view domain while preserving particle morphology, enabling reliable segmentation and tracking during rapid granular flows.

**APPENDIX B: EDWARDS STATISTICAL FRAMEWORK IN THE FLUID REGIME**

For static packings, the Edwards ensemble describes fluctuations of a coarse-grained volume $V$ by a Boltzmann-like distribution, $P(V) = \frac{\Omega(V)}{Z(\chi)} \exp(-V/\chi)$, where $\chi$ is the compactivity (a temperature-like variable), $\Omega(V)$ is the density of states, and $Z(\chi)$ is the partition function [54]. The coarse-grained volume $V$ was calculated by summing the Voronoi volumes of the 15 nearest neighbors surrounding each particle, $V = \sum_{n=1}^{15} V_{\text{voro}}$.

The compactivity $\chi$ is computed using a histogram overlapping method, which compares

the volume distribution $P(V)$ of a given state with that of a reference state $r$:

$$\frac{P^r(V)}{P(V)} = \frac{Z(\chi)}{Z(\chi^r)} \exp\left[\left(\frac{1}{\chi} - \frac{1}{\chi^r}\right)V\right]. \tag{B1}$$

For static packings, the reference state was chosen as the random loose packing (RLP) state, which represents the loosest mechanically stable packing attainable and is defined to have infinite compactivity (i.e., $\chi^r \to \infty$). $\chi$ was further converted into an energy-scale Edwards temperature $T_{\text{EDW}} = P\chi$, enabling direct comparison with the kinetic temperature $T_k$.

In flowing states, strict mechanical stability is violated owing to frequent particle collisions, as evidenced experimentally by the significantly reduced contact number $Z$ in flowing states compared with static packings at the same $\phi$, see Supplemental Material Fig. S4(b) [25]. Consequently, mechanically unstable configurations become accessible, and the phase space traversed by granular flows differs from that of static packings. This change is encoded in a modified density of states, $\Omega_f(V)$, and the effective temperature in flowing granular states is therefore expected to incorporate contributions from both granular kinetic motion and configurational fluctuations. Importantly, as shown in the Supplemental Material Fig. S4(a) [25], despite the absence of strict mechanical stability, the volume distribution in flowing states retains a Boltzmann-like form, $P_f(V) = \frac{\Omega_f(V)}{Z(T_{\text{eff}})} \exp\left(-\frac{pV}{T_{\text{eff}}}\right)$. For two flowing states under different driving conditions, as shown in the inset of Supplemental Material Fig. S4(a) [25], the histogram overlapping method yields a linear relation, , $ln\left[\frac{P_f^r(V)}{P_f(V)}\right] \propto \left(\frac{1}{T_{\text{eff}}} - \frac{1}{T_{\text{eff}}^r}\right)$, demonstrating that the density of states $\Omega_{\text{flow}}(V)$ remains invariant across flowing states and therefore cancels out. This invariance indicates that, despite being driven and out of equilibrium, the system maintains statistical ergodicity within the flowing regime. This histogram overlapping method thus both validates the persistence of a well-defined ensemble and enables a consistent definition of $T_{\text{eff}}$ in flowing states, with the reference effective temperature $T_{\text{eff}}^r$ accounting for kinetic contributions. In the dilute regime below $\phi_{\text{RLP}}$, configurational relaxation and short-time dynamics are not separated, such that $T_{\text{eff}}$ is identical to the kinetic temperature $T_k$. Accordingly, the RLP state was assigned a finite effective temperature equal to its granular kinetic temperature, $T_{\text{eff}}^r = T_{k,\text{RLP}} = 0.036 P d^3$, which serves as a natural reference point for extending the Edwards framework to flowing granular states.

## APPENDIX C: DIFFUSION OF THE EFFECTIVE TEMPERATURE

We attribute sub-yield creep to the spatial diffusion of the effective temperature $T_{\text{eff}}$. Owing to the azimuthal symmetry of the Couette shear cell, $T_{\text{eff}}$ depends only on the radial coordinate $R$ and satisfies

$$D_T d^2 \nabla^2 T_{\text{eff}}(R) = T_{\text{eff}}(R) - T_{\text{eff,loc}}(R), \tag{C1}$$

where $D_T$ denotes the thermal diffusivity and $T_{\text{eff,loc}}$ represents the local generation of effective temperature by shear in the absence of diffusion. To evaluate $\nabla^2 T_{\text{eff}}(R)$ from experimental data, the measured profile was fitted with a smooth functional form, $T_{\text{eff}}(R) = T_0 \exp(\sum_{n=0}^{3} a_n R^n)$, and the Laplacian, $\nabla^2 T_{\text{eff}} = \frac{1}{R} \frac{d}{dR}\left(R \frac{dT_{\text{eff}}}{dR}\right)$, was computed analytically from the fitted function. The local source term is constrained by the no-diffusion condition $\nabla^2 T_{\text{eff,loc}}(R) = 0$, which under radial symmetry yields $T_{\text{eff,loc}}(R) = C_0 \ln\left(\frac{R}{R_0}\right) + C_1$, where $C_0$ and $C_1$ are constants. At the confining walls, $T_{\text{eff,loc}}$ is set equal to the experimentally measured $T_{\text{eff}}$. Because no local flow occurs once $\phi$ exceeds $\phi_{\text{RLP}}$, $T_{\text{eff,loc}}$ was additionally constrained to a constant value, $T_{\text{eff,loc}}(R) = C_1$, in the regions where $\phi > \phi_{\text{RLP}}$ [blue curve in the inset of Fig. 5(b)]. With $T_{\text{eff}}(R)$, $\nabla^2 T_{\text{eff}}(R)$, and $T_{\text{eff,loc}}(R)$ specified, Equation (C1) yields the spatial profile of $D_T$, which remains approximately constant ($D_T \approx 3.5$) throughout the dense regime. By contrast, replacing $T_{\text{eff}}$ with the kinetic temperature $T_k$ in Eq. (C1) leads to a rapidly increasing effective thermal diffusivity upon densification [red curve in the inset of Fig. 5(b)], with $D_T$ rising from approximately 3.5 at $\phi_{\text{RLP}} = 0.59$ to above 30 at $\phi = 0.63$.

### APPENDIX D: SPATIAL CORRELATION OF PARTICLE DYNAMICS

The spatial cooperativity of particle dynamics was quantified using a velocity correlation function $C_v$ within each concentric ring. For particles $i$ and $j$ separated by a distance $r_{ij}$, we calculate $C_v(r,R)$ by:

$$C_v(r,R) = \frac{\langle \sum_{i,j} v_i v_j \cdot \delta(r_{ij} - r) \rangle_R}{\langle \sum_{i,j} \delta(r_{ij} - r) \rangle_R}, \tag{D1}$$

where $v_i$ is the vertical velocity of particle $i$, and $\langle \cdots \rangle_R$ denotes averaging over all particle pairs within the ring at radial position $R$ [59]. As shown in Supplemental Material Fig. S5 [25], $C_v(r)$ decays with increasing $r$ and exhibits oscillations whose peak positions coincide with the maxima of the pair correlation function $g(r)$, reflecting the underlying packing structure. The amplitudes of the first four peaks of $C_v(r)$ were fitted with an exponential function, $C_v(r) \propto \exp(-r/\xi_C)$, and the correlation length $\xi_C$ reported in the main text was extracted from the fitting. Additionally, at lower driving rates, the peak in $\xi_C$ shifts toward smaller $\phi$, signaling that boundary confinement truncates mesoscopic cooperativity and suppresses collective rearrangements, thereby enhancing resistance to flow.

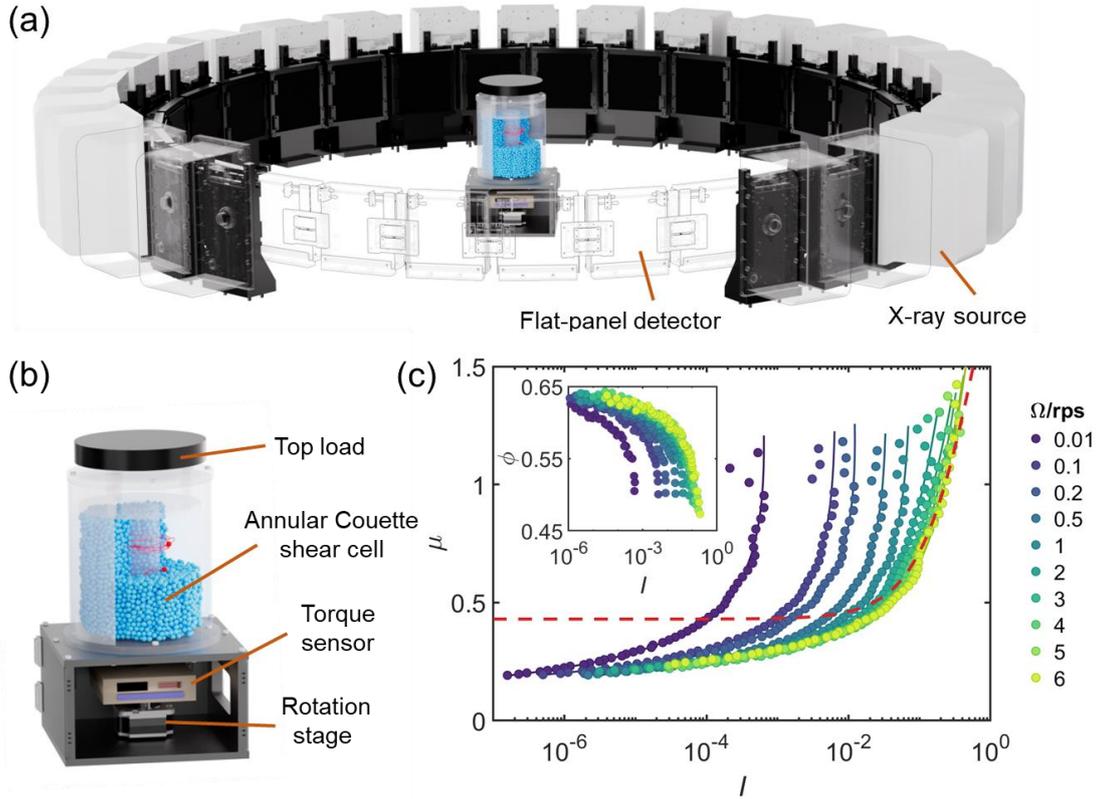

FIG. 1. (a) Schematic of the multi-source X-ray tomographic system. (b) Schematic of the annular Couette shear cell, superimposed by several representative particle trajectories under shear (red curves). (c) Macroscopic friction coefficient $\mu$ as a function of the inertial number $I$ at different rotation rates $\Omega$. The red dashed curve shows the empirical $\mu(I)$ relation, $\mu(I) = \mu_s + \frac{\mu_m - \mu_s}{1 + I_0/I}$, with $\mu_s = 0.43$, $\mu_m = 3.12$, and $I_0 = 0.012$; solid curves correspond to $\mu(I)$ profiles derived from fitted velocity fields (see *Methods*). Inset: Packing fraction $\phi$ as a function of $I$ for different $\Omega$. All data correspond to SS systems.

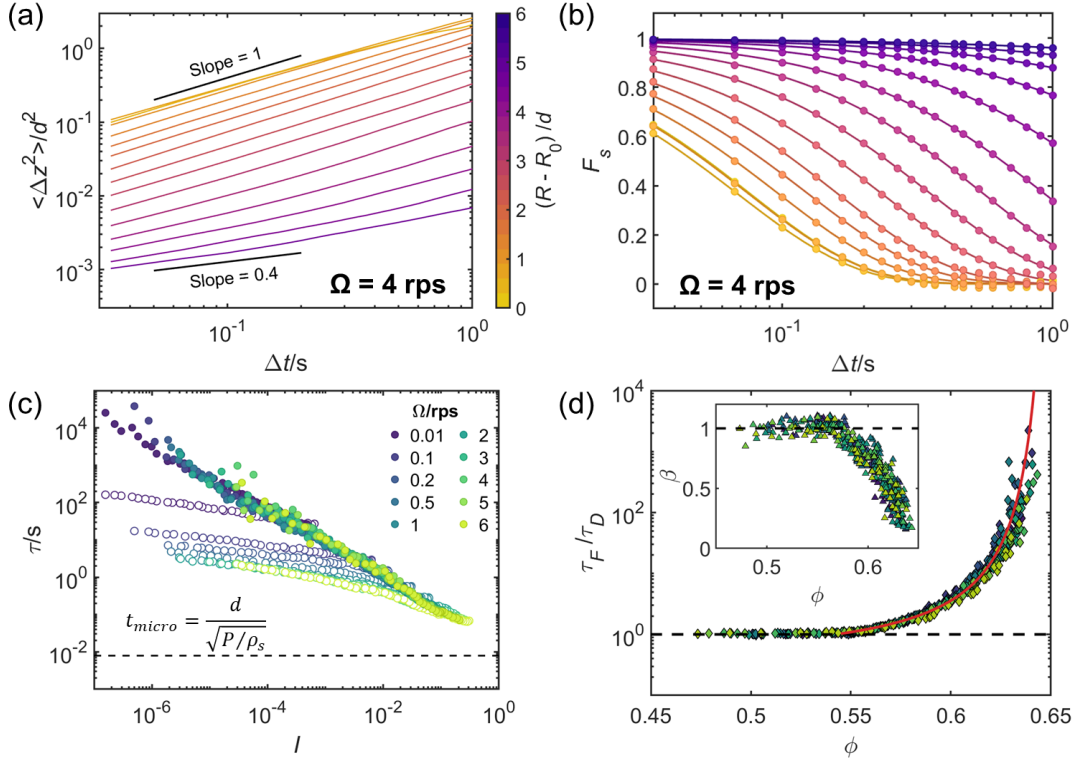

FIG. 2. (a) MSDs along the vertical direction for different radial positions $R$ at $\Omega = 4$ rps. The short-time diffusion exponent decreases from 1 near the shearing inner wall to approximately 0.6 at larger $R$, signaling a crossover from collision-dominated to contact-dominated flow regimes. (b) Self-intermediate scattering function $F_s(\Delta t)$ for different $R$ at $\Omega = 4$ rps. Solid curves show KWW fits. (c) Diffusive timescale $\tau_D$ (open symbols) and structural relaxation time $\tau_F$ (filled symbols) as functions of the inertial number $I$ for systems driven at different $\Omega$. The dashed line denotes the estimated microscopic timescale $t_{micro}$ of the $\mu(I)$ framework. (d) Ratio of the structural relaxation time to the diffusive timescale, $\tau_F/\tau_D$, as a function of the volume fraction $\phi$. The black dashed line marks $\tau_F = \tau_D$. The red curve shows a VFT-like growth, $\tau_F/\tau_D \propto \exp[A/(\phi_c - \phi)]$, with $\phi_c = 0.648 \pm 0.011$. Inset: Stretching exponent $\beta$ extracted from the KWW fits; the black dashed line corresponds to $\beta = 1$. All data correspond to SS systems.

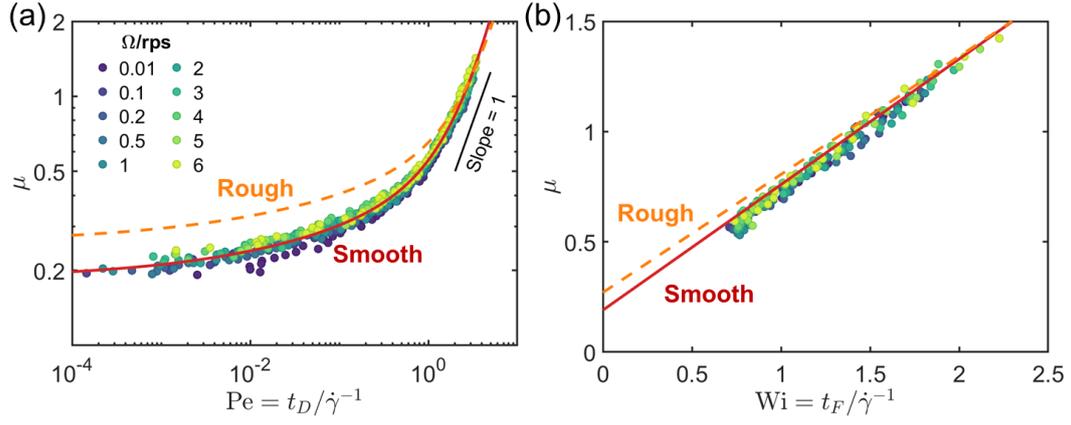

FIG. 3. (a) Macroscopic friction coefficient $\mu$ as a function of the Péclet number $\text{Pe} = \tau_D/\dot\gamma^{-1}$ for different driving rates $\Omega$. (b) $\mu$ plotted against the Weissenberg number $\text{Wi} = \tau_F/\dot\gamma^{-1}$. Red solid and orange dashed curves indicate representative trends for SS and RS systems, respectively. In (a), the data follow a Herschel-Bulkley form, $\mu = \mu_c + K\text{Pe}^{0.3}$, whereas in (b) $\mu$ exhibits an approximately linear dependence on Wi.

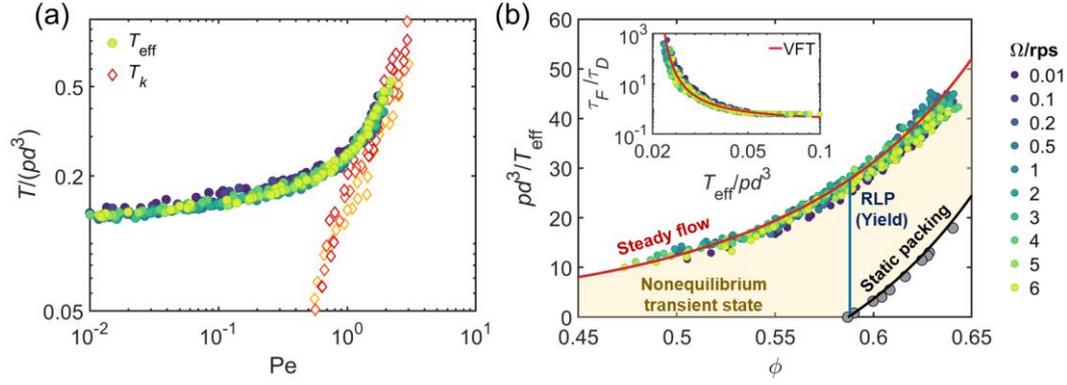

FIG. 4. (a) Effective temperature $T_{\text{eff}}$ (solid circles) and kinetic temperature $T_k$ (open diamonds) as functions of the Péclet number Pe. (b) Equation of state relating $T_{\text{eff}}$ to packing fraction $\phi$ for steady granular flows (colored symbols) and static packings (grey symbols; data from ref. [45]). The red curve shows the Carnahan-Starling equation of state for hard spheres. Shaded regions denote non-equilibrium transient states; blue solid line marks the RLP density, corresponding to the yield point. Inset: Ratio $\tau_F/\tau_D$ as a function of $T_{\text{eff}}$ for different $\Omega$. The red curve shows a VFT-like divergence, $\tau_F/\tau_D \propto \exp[A_T/(T_{\text{eff}} - T_c)]$, with $T_c = 0.045 P d^3$. All data correspond to SS systems.

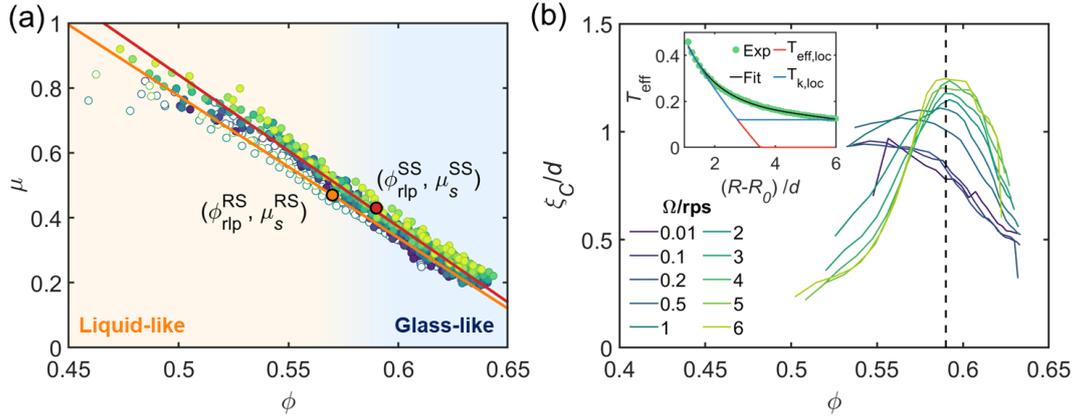

FIG. 5. (a) Relation between the macroscopic friction coefficient $\mu$ and the packing fraction $\phi$ for SS (filled symbols) and RS (open symbols) systems. Solid lines indicate linear fits of the form $(\mu - \mu_s) \propto (\phi_{\text{RLP}} - \phi)$. Filled markers denote the RLP volume fraction $\phi_{\text{RLP}}$, which coincides with the static yield friction $\mu_s$ for each particle type. The shaded background distinguishes liquid-like ($\phi < \phi_{\text{RLP}}$) and glass-like ($\phi > \phi_{\text{RLP}}$) regimes. The location of $\phi_{\text{RLP}}$, and hence the yield boundary, depends on interparticle friction. (b) Spatial correlation length $\xi_C$ of particle motion as a function of $\phi$ for SS systems. The dashed line indicates $\phi_{\text{RLP}}$. Inset: Radial profile of the effective temperature $T_{\text{eff}}$. Experimental data (symbols) are well described by a diffusive model of $T_{\text{eff}}$ with a constant thermal diffusivity (blue curves). By contrast, nonlocal $\mu(I)$ models based on the kinetic temperature $T_k$ (red curve) decay too rapidly and reproduce the experimental trend only by invoking a diverging thermal diffusivity.

# Supplemental Material for

# A Unified Glassy Rheology for Granular Matter

Zhikun Zeng[1,*], Jiazhao Xu[1,*], Hanyu Li[2,*], Shiang Zhang[3], Houfei Yuan[1], Chijin Zhou[2], Xueliang Dai[2], Haiyang Lu[1], Xin Wang[1], Jun Zhao[3], Yonglun Jiang[2], Zhuan Ge[2], Gang Huang[2], Chengjie Xia[4], Jianqi Sun[3,†], Yan Xi[5,‡], and Yujie Wang[1,2,6,§]

[1]*School of Physics and Astronomy, Shanghai Jiao Tong University, Shanghai, China.*
[2]*College of Physics, Chengdu University of Technology, Chengdu, China.*
[3]*Med-X Research Institute and School of Biomedical Engineering, Shanghai Jiao Tong University, Shanghai, China.*
[4]*School of Physics, East China Normal University, Shanghai, China.*
[5]*Jiangsu First-Imaging Medical Equipment Co., Ltd., Jiangsu, China.*
[6]*State Key Laboratory of Geohazard Prevention and Geoenvironment Protection, Chengdu University of Technology, Chengdu, China.*

[*]These authors contributed equally to this work.

[†]Contact author: milesun@sjtu.edu.cn

[‡]Contact author: yanxi@first-imaging.com

[§]Contact author: yujiewang@sjtu.edu.cn

# I. SUPPLEMENTARY FIGURES

## A. Architecture of the S-STAR Net

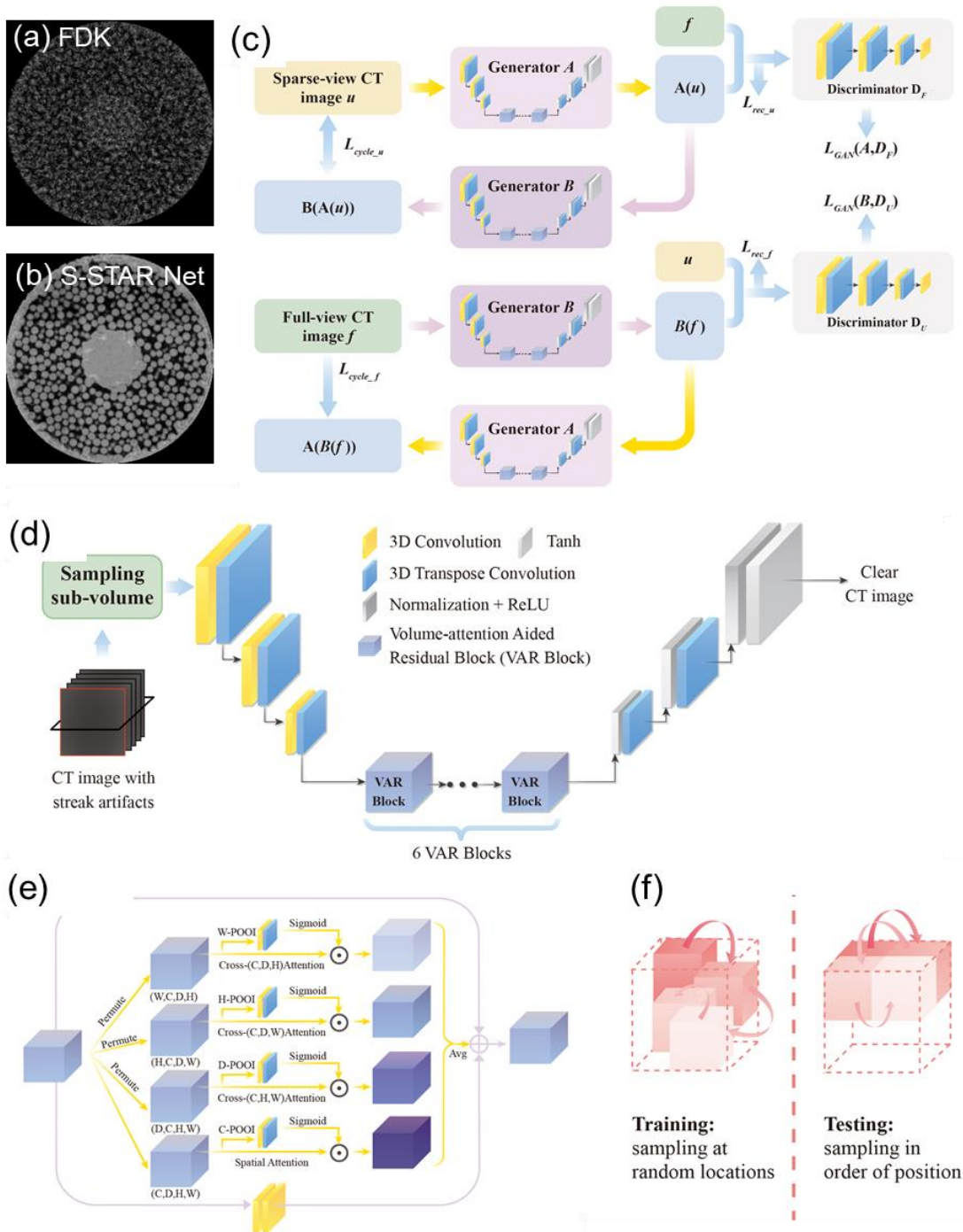

FIG. S1. (a,b) Representative cross-sectional slices reconstructed from the same sparse-view projections using (a) the conventional FDK algorithm and (b) the S-STAR Net. Pronounced streak artifacts in the FDK reconstruction are substantially suppressed by S-STAR Net. (c) Overall architecture of the S-STAR Net. Artifact reduction is formulated as a domain translation

from the ultra-sparse-view domain ($U$) to the full-view domain ($F$). The network is trained using a cyclic adversarial framework. Generators $A$ and $B$ implement the mappings $U \rightarrow F$ and $F \rightarrow U$, respectively, with corresponding discriminators $D_F$ and $D_U$. $L_{GAN}(A, D_F)$ and $L_{GAN}(B, D_U)$ denote the adversarial losses; $L_{DE}$ and $L_{rec}$ denote the domain-enhancement and reconstruction losses, respectively. $u$ and $f$ represent paired samples from $U$ and $F$. (d) Architecture of the generators. (e) Structure of the volume-attention aided residual (VAR) block. $C$, $D$, $H$, and $W$ denote the channel, depth, height, and width dimensions, respectively. (f) Sliding sub-volume sampling strategy used during training and inference.

## B. Flow geometry

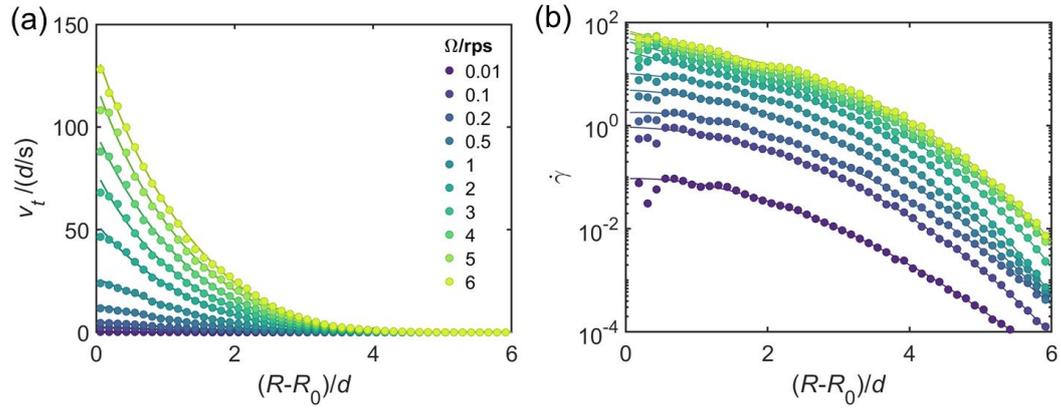

FIG. S2. (a) Mean azimuthal velocity $v_t$ as a function of the radial distance $R - R_0$ from the inner wall at different driving rates $\Omega$. (b) Corresponding shear-rate profiles $\dot{\gamma}(R - R_0)$ for the same conditions. All data correspond to SS systems.

## C. Representative particle trajectories

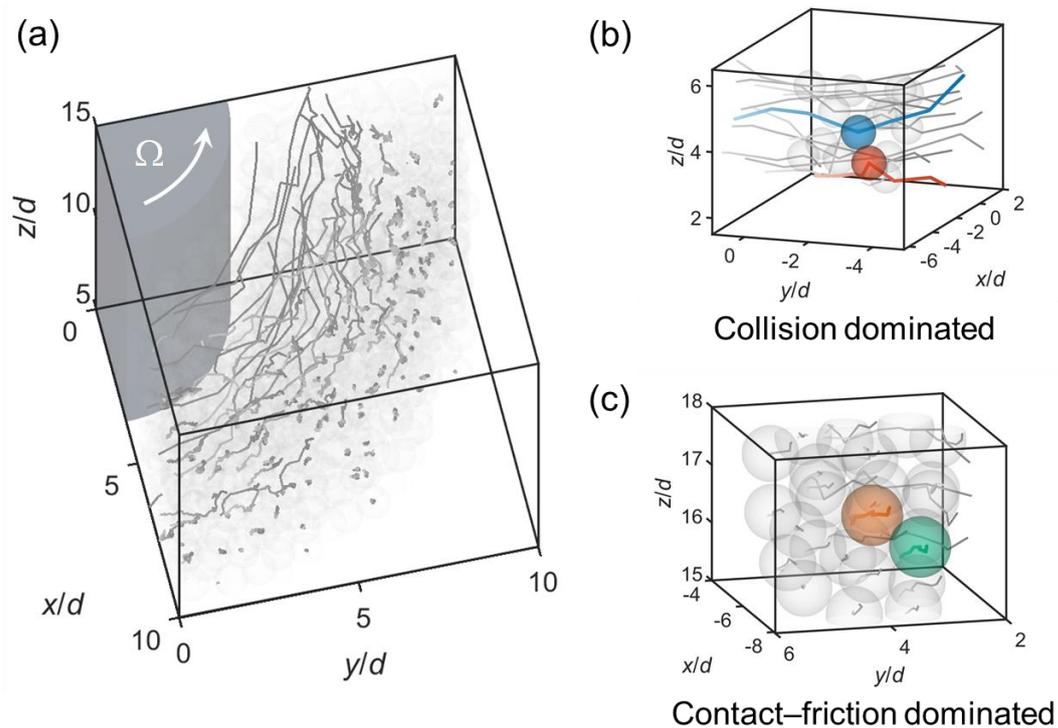

FIG. S3. (a) Trajectories of selected particles under Couette shear. The grey central cylinder represents the rotating inner wall, driven at a rate $\Omega$. (b,c) Representative particle trajectories in the collision-dominated (b) and contact-friction-dominated (c) regimes, respectively. Curve color (light to dark) indicates temporal progression. All data correspond to SS systems at $\Omega = 2$ rps.

## D. Effective temperature from volume statistics and contact number

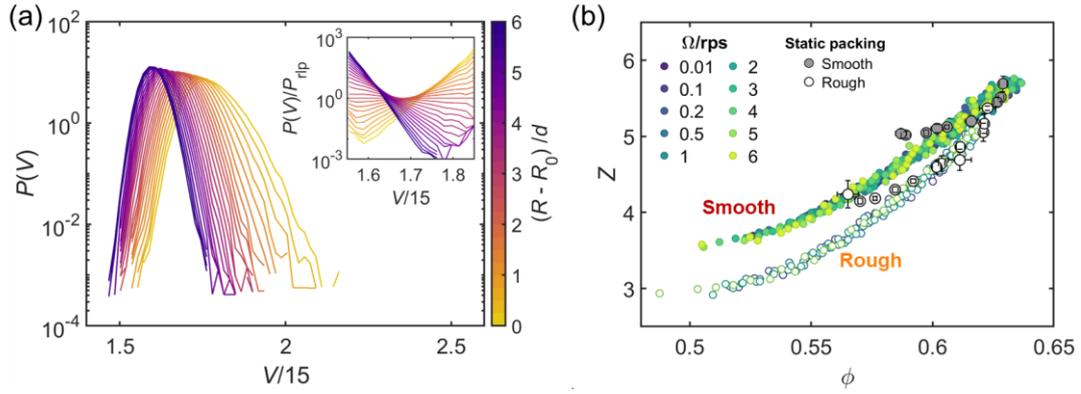

FIG. S4. (a) Volume distributions $P(V)$ at different radial positions $R$ for the SS systems with $\Omega = 2$ rps. Inset: Ratio of $P(V)$ in sheared granular flows to that in the RLP state. (b) Contact number Z as a function of the packing fraction $\phi$ for SS (filled symbols) and RS (open symbols) systems. Colored symbols correspond to steady sheared flows at different driving rates $\Omega$, and grey symbols denote static packings prepared by tapping (data from ref. [45]).

## E. Spatial structural and velocity correlations

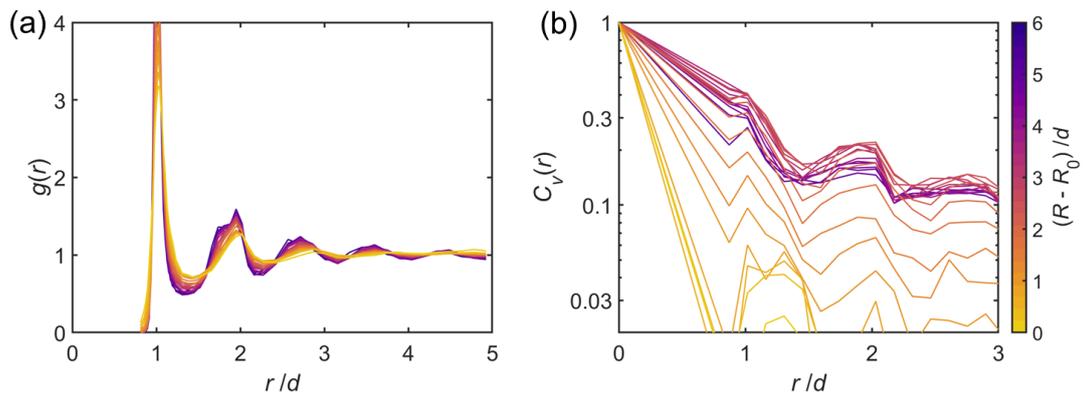

FIG. S5. (a) Pair correlation function $g(r)$ as a function of the interparticle distance $r$. (b) Spatial displacement correlation $C_v(r)$ as a function of $r$. All data correspond to SS systems at $\Omega = 2$ rps.

## II. VIDEO CAPTIONS

**Video 1.** Particle dynamics under annular Couette shear at $\Omega = 4$ rps.